\documentclass[usenatbib]{iaus}
\usepackage{graphicx}

\title[CMD for candidate agegap filling LMC clusters] 
{Colour-Magnitude Diagrams of candidate age-gap filling LMC clusters}

\author[Eduardo Balbinot, Bas\'{\i}lio Santiago, Leandro Kerber {\it et al.}]
{Eduardo Balbinot$^1$, Bas\'{\i}lio X. Santiago$^1$, Leandro Kerber$^2$, Beatriz Barbuy$^2$ 
\and Bruno M. S. Dias$^2$}

\affiliation{
$^1$ Departamento de Astronomia, Universidade Federal do Rio Grande do Sul \\ Porto Alegre, RS, Brazil \\ email: {\tt (balbinot, santiago)@if.ufrgs.br}
\\[\affilskip]
$^2$ IAG, Universidade de Sao Paulo \\ Sao Paulo, SP, Brazil }

\pubyear{2008}
\volume{256}  
\pagerange{001--003}
\jname{The Magellanic System: Stars, Gas, and Galaxies}
\editors{Jacco Th. van Loon \& Joana M. Oliveira, eds.}
\begin{document}

\maketitle

\begin{abstract}
The LMC has a rich star cluster system spanning a wide range of ages and 
masses. One striking feature of the LMC cluster system is the existence 
of an age gap between 3-10 Gyrs. Four LMC clusters whose integrated colours 
are consistent with those of intermediate age simple stellar populations 
have been imaged with the Optical Imager (SOI) at the Southern Telescope 
for Astrophysical Research (SOAR). Their colour-magnitude diagrams (CMDs) 
reach V $\sim$ 24. Isochrone fits, based on Padova evolutionary models, were 
carried out to these CMDs, after subtraction of field contamination. The 
preliminary results are as follows: KMK88-38 has an age of about 1.3 Gyr, 
assuming typical LMC metallicity and distance modulus, and a very low 
redenning. For OGLE-LMC0531, the best eye fits to isochrones yield an 
age $\sim$ 1.6 Gyr and E(B-V)=0.03. BSDL917 is younger, $\sim$ 150 Myrs, and 
subjected to larger extinction (E(B-V)=0.08). The remaining cluster is 
currently under analysis. Therefore, we conclude that these clusers are 
unlikely to fill in the LMC cluster age gap, even when fitting uncertainties 
in the parameters are considered.

\end{abstract}

\firstsection 
              
\section{Introduction}

The LMC has a rich star cluster system spanning a wide range of ages and masses. 
One striking feature of the LMC cluster system is the existence of an age gap 
between 3-10 Gyrs. Four LMC clusters whose integrated colours are consistent with 
those of intermediate age simple stellar populations have been imaged with the 
Optical Imager (SOI) at the Southern Telescope for Astrophysical Research (SOAR).

\section{Data}

We have imaged 3 out of the 5 LMC clusters listed by Hunter et al (2003) as belonging to the age gap. Two of them have been fully reduced.

The images were taken in 2007 with SOAR/SOI telescope in the B, V, and I filters. A slow readout was used in order to minimise detector noise. A 2x2 pixel binning was adopted, yielding a spatial scale of 0.154 arcsec/pixel. Seeing was always around 0.8 arcsec.

The exposures were flatfielded, bias subtracted, mosaiched and latter combined. The photometry was carried out with standard point spread function (PSF) fitting. The DAOPHOT package (Stetson 1994) was used to detect sources (4 $\sigma$ above sky background) and measure magnitudes. The PSF was modelled as a Moffat function with $\beta=1.5$. Photometric calibration was obtained from the standard fields in the Northeast arm of the SMC (Sharpee et al. 2002). A typical SOAR/SOI image section is shown in Figure 1.

\begin{figure}
\centering
\includegraphics[width=0.65\textwidth]{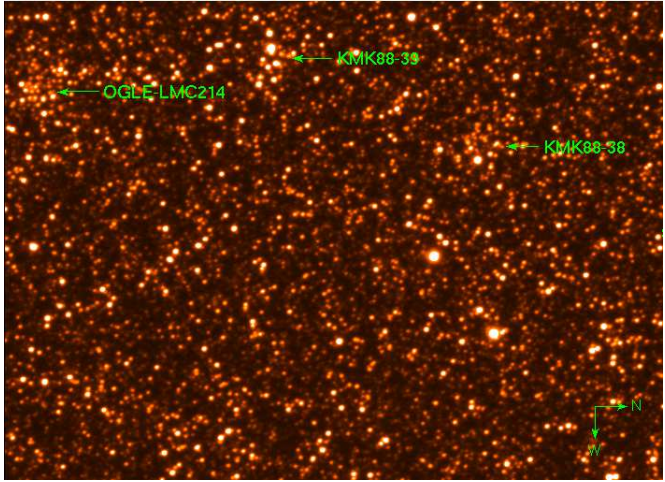} 
\caption{A 2.6' x 3.6' image section of the field around the cluster KMK88-38,
where two other known clusters are included. Their positions in the image are indicated.}
\label{}
\end{figure}

CMDs for the fields of the two age gap candidates already reduced are shown in Figure 2. Their colour-magnitude diagrams (CMDs) reach $V \sim 23$. From left to right, we show the full SOI field CMD, the CMD in the cluster region and the field subtracted cluster CMD. Padova isochrones from Girardi et al. (2002) are superposed to these latter. Field subtraction was performed statistically, applying the method described in detail by Kerber et al. (2002).

\begin{figure}
\centering
\includegraphics[width=0.55\textwidth]{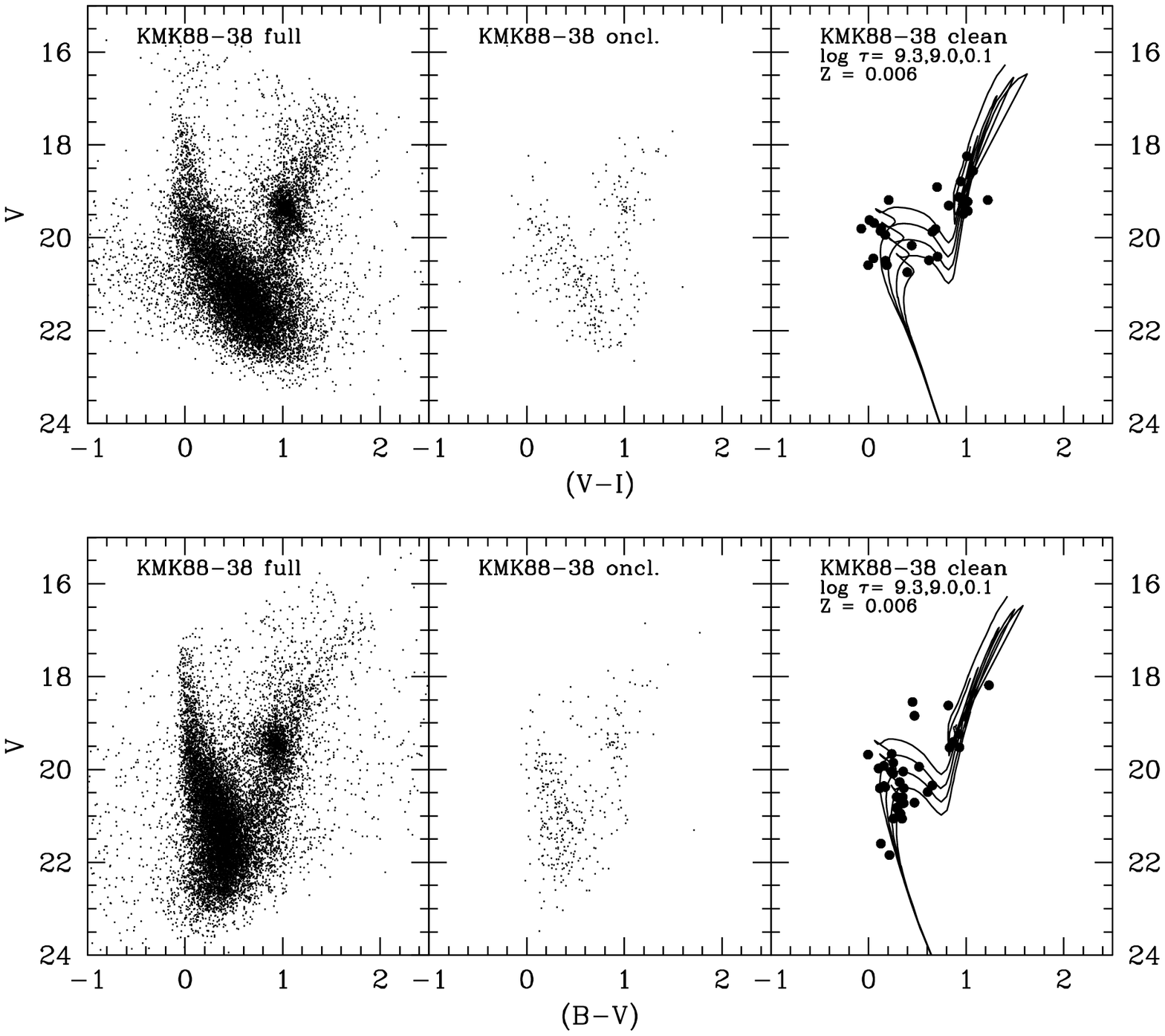} \\
\includegraphics[width=0.55\textwidth]{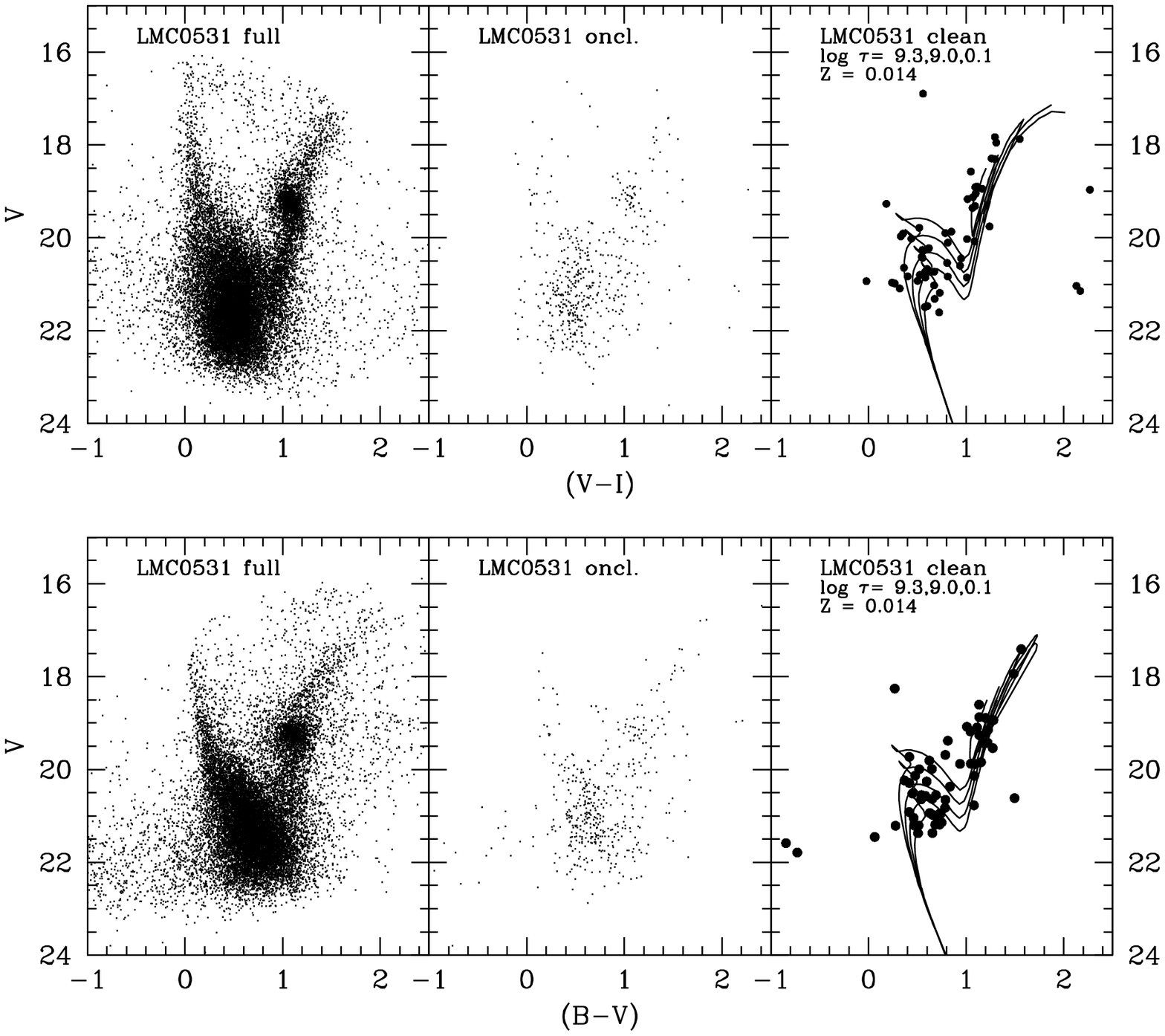} 
\caption{V,(V-I) (top) and V,(B-V) (bottom) CMDs for
the fields around KMK88-38 (left) and LMC0531 (right) clusters. The
panels from left to right show: the entire SOI field, the region around
the cluster, the result of field subtraction. Padova isochrones are overlaid
to the latter panels. Ages are shown as $log(\tau_{max}(yrs))$, $log(\tau_{min})$,
$\Delta log \tau$. The adopted metallicity is also shown.}
\label{}
\end{figure}

Besides the age gap candidates, the SOI images also covered
other LMC clusters listed in the catalogue by Bica et al. (2008). Figure 3
shows the field subtracted CMDs for them, again with isochrones overlaid. 

\begin{figure}
\centering
\includegraphics[width=0.45\textwidth]{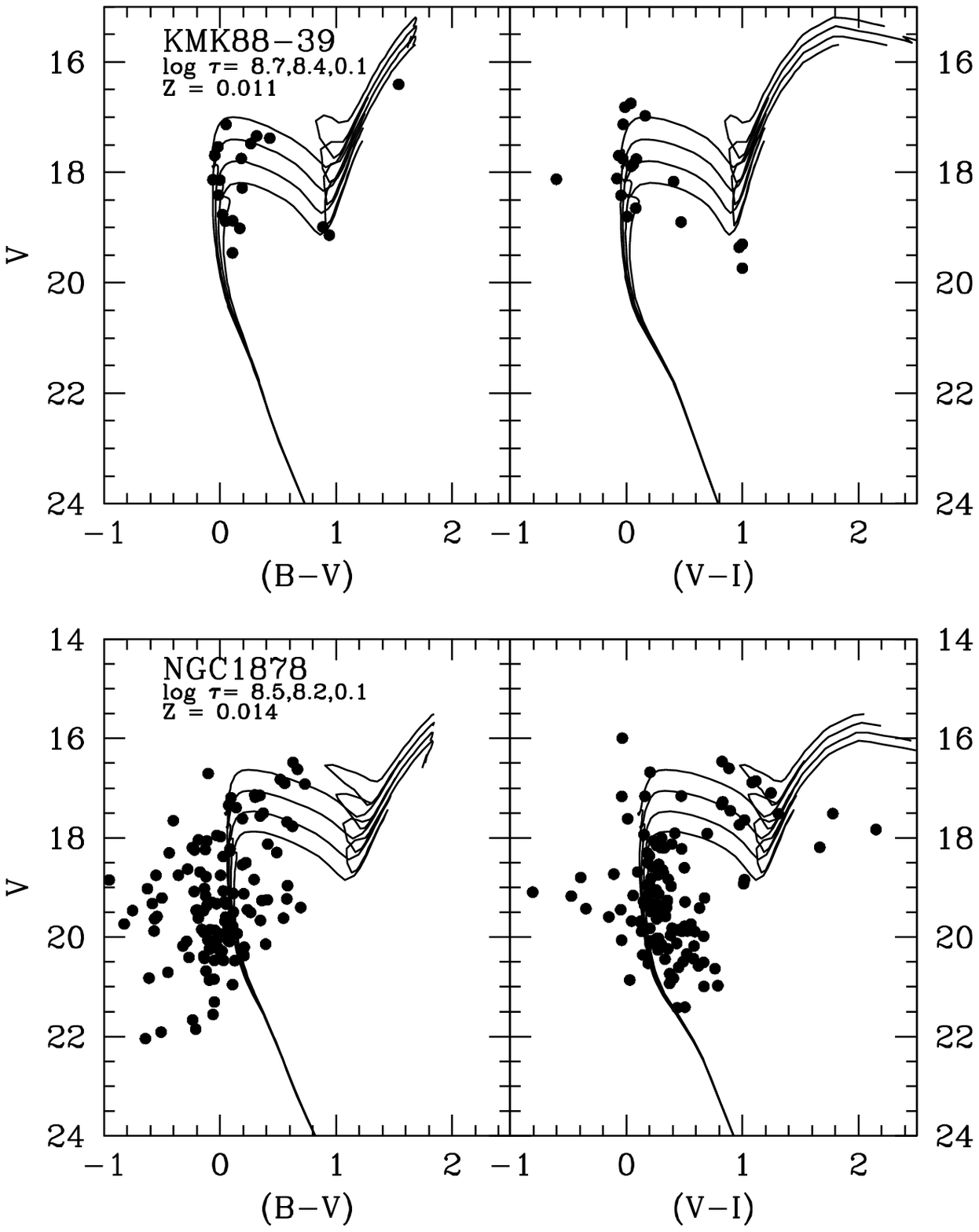} 
\includegraphics[width=0.45\textwidth]{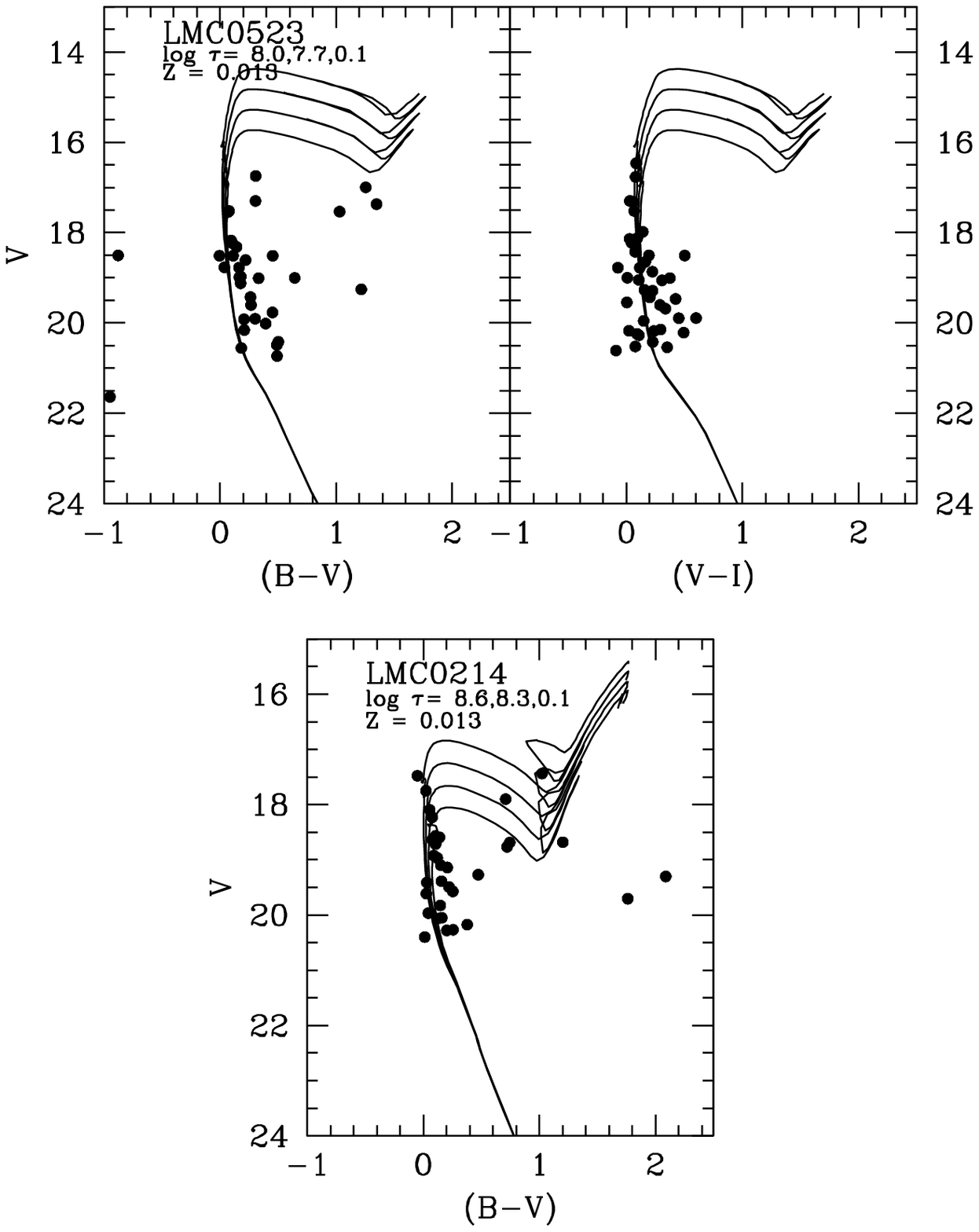}
\caption{Field subtracted CMDs for the remaining
clusters found in our SOAR/SOI images. The symbols are the same
as in the corresponding panels of Figure 2.}
\label{}
\end{figure}

Visual matching of the observed CMDs to the isochrone set has allowed
us to estimate age, metallicity, distance modulus and reddening for each
cluster. The resulting parameters are shown in Table 1. 

\section{Results and conclusions}

The original targets, KMK88-38 and LMC0531, turn out to be the relatively
old, as expected, with ages $\sim 1-2$ Gyrs. However, they are still younger
than the lower age limit of the LMC gap. Interestingly, KMK88-39, LMC0214
and LMC0523, which lie in the same SOI fields (5.5arcmin x 5.5 arcmin in
size), are much younger. LMC0523 final V,(B-V) CMD has 3 stars in the
Red Clump region. Even though they survived field subtraction, these stars
have relatively low probabilities of actually belonging to the cluster,
and they are, in fact, absent from the V,(V-I) CMD. For LMC0214 we have
only B and V images. The ages for this latter, as well as for KMK88-39
are upper limits, as their upper main sequence is close to the
saturation limit. Finally, NGC 1878 is a richer and denser cluster. Field
subtraction was not as efficient in this case, especially in V,(B-V). We
attribute that to crowding effects, which make photometric errors larger
in the cluster region than in the field, jeopardising the statistical field
subtraction. Still, its V,(V-I) CMD indicates that NGC 1878
is also younger than 0.5 Gyr.

\begin{table}
\centering
\begin{tabular}{c|c|c|c|c}

$ Name $&$ log(Age) $&$ Z $&$ E(B-V) $& $(m-M)_V $\\
	\hline OGLE-LMC0214 &$ 8.4 \pm 0.3 $&$ 0.013 $&$ 0.10 $ &$ 18.50 $\\
	\hline OGLE-LMC0523 &$ 7.8 \pm 0.3 $&$ 0.013 $&$ 0.20 $ &$ 18.40 $\\ 
	\hline OGLE-LMC0531 &$ 9.2 \pm 0.2 $&$ 0.014 $&$ 0.09 $ &$ 18.50 $\\ 
	\hline KMK88-38     &$ 9.2 \pm 0.2 $&$ 0.006 $&$ 0.01 $ &$ 18.65 $\\ 
	\hline KMK88-39     &$ 8.5 \pm 0.3 $&$ 0.011 $&$ 0.02 $ &$ 18.50 $\\ 
	\hline NGC1878      &$ 8.4 \pm 0.3 $&$ 0.014 $&$ 0.17 $ &$ 18.50 $\\ 
	\hline 

\end{tabular} 
\caption{The parameters found for our sample.}
\end{table}

We thus conclude that the sample analysed so far does not contain any
cluster located in the LMC age gap. We are currently reducing the field
images of another age gap candidate: LMC0169. We are also reducing lower
exposure time images of all clusters and perfecting the field subtraction
algorithm; both are important steps towards improving our age and metallicity
constraints on the clusters.

\end{document}